\documentclass[12pt]{article}

\usepackage{hyperref}
\usepackage{amssymb}
\usepackage[dvips]{graphicx}

\begin{document}

\vspace{3cm}
\begin{center}
{\Large{\bf Affleck-Dine leptogenesis with triplet Higgs}}

\vspace{1cm}

{\large{
Masato Senami\footnote{E-mail address: senami@nucleng.kyoto-u.ac.jp}
and
Katsuji Yamamoto\footnote{E-mail address: yamamoto@nucleng.kyoto-u.ac.jp}
}}

\vspace{1cm}

{\it Department of Nuclear Engineering, Kyoto University,
Kyoto 606-8501, Japan}

\vspace{1cm}
{\bf Abstract}
\end{center}

\bigskip

We study an extension of the supersymmetric standard model
including a pair of electroweak triplet Higgs
$ \Delta $ and $ {\bar \Delta} $.
The neutrinos acquire Majorana masses mediated by these triplet Higgs fields
rather than the right-handed neutrinos.
The successful leptogenesis for baryogenesis can be realized
after the inflation through the Affleck-Dine mechanism
on a flat manifold consisting of $ \Delta $, $ {\bar \Delta} $,
$ {\tilde e}^c $ (anti-slepton),
even if the triplet Higgs mass $ M_\Delta $ is much larger
than the gravitino mass $ m_{3/2} \sim 10^3 {\rm GeV} $.
Specifically, due to the effects of the potential terms
provided with the superpotential terms $ M_\Delta {\bar \Delta} \Delta $,
$ ( \lambda_{L \! \! \! /} / 2M ) {\bar \Delta} {\bar \Delta} e^c e^c $,
$ ( \lambda_\Delta / 2M ) {\bar \Delta} \Delta {\bar \Delta} \Delta $
($ \lambda_{L \! \! \! /} / \lambda_\Delta \sim 0.3 - 3 $),
the phases of $ \Delta $, $ {\bar \Delta} $, $ {\tilde e}^c $ are rotated
at the time with the Hubble parameter $ H \sim M_\Delta $,
producing generally the asymmetry with fraction $ \epsilon_L \sim 0.1 $.
If $ M_\Delta $ is large enough, this early leptogenesis can be completed
before the thermal effects take place.

\newpage

\section*{Introduction}

The experiment of the atmospheric neutrinos
by the SuperKamiokande collaboration
indicated a convincing evidence for the neutrino masses and oscillations
\cite{fukuda}.
This hence will require some extension of the standard model
by introducing lepton number nonconserving interactions.
The natural explanation of the small Majorana neutrino masses
is usually made by the see-saw mechanism
with heavy right-handed electroweak singlet neutrinos
\cite {seesaw}.
It is also noted that the effective higher dimensional operator
$ L L H_u H_u $ providing the small neutrino masses
is generated even through the exchange
of heavy electroweak triplet Higgs fields
($ L $ and $ H_u $ are the lepton and Higgs doublets, respectively)
\cite{triseesaw}.

The lepton number violating interactions,
which may be provided with the right-handed neutrinos
or the triplet Higgs as mentioned above for the neutrino masses,
will even have important effects in the early universe.
In particular, they can be relevant for the generation
of lepton number asymmetry, i.e., leptogenesis.
Then, the lepton number asymmetry can further produce
the sufficient baryon number asymmetry
through the anomalous sphaleron processes in the electroweak gauge theory
\cite{fuya}.
Therefore, the leptogenesis in the early universe
is a very interesting issue in the extensions of the standard model
involving the lepton number violating interactions.

There are two familiar types of scenarios for leptogenesis.
One is due to the non-equilibrium decays of heavy particles,
and the other is the Affleck-Dine mechanism
\cite{AD,DRT}.
Both scenarios have been explored extensively in the ordinary see-saw case,
i.e., the decay of right-handed neutrinos
which are produced either thermally or non-thermally
\cite{fuya,ndecay,ahky},
and the Affleck-Dine mechanism with the $ {\tilde L} H_u$ flat direction
\cite{DRT,muya,nad,hay}.
In this Affleck-Dine leptogenesis scenario,
the lepton number asymmetry and the neutrino masses are related directly.
It is clarified recently that if the thermal effects
\cite{ACE}
and gravitino problem
\cite{rehe,ntgra}
are taken into account, the mass of the lightest neutrino should be less
than of the order of $10^{-8}$ eV
\cite{hay}
to generate the sufficient lepton number asymmetry.

On the other hand, the leptogenesis in the triplet Higgs case
has not been examined fully so far.
Recently, the non-equilibrium decay of triplet Higgs
has been considered in a supersymmetric model
\cite{masa,trithe}.
In this triplet Higgs decay scenario,
it is required that the mass of triplet Higgs should be less
than the reheating temperature $ T_R $ after inflation,
which is bounded to be lower than $ 10^8 {\rm GeV} - 10^{10} {\rm GeV} $
to avoid the gravitino problem
\cite{rehe,ntgra}.
It is also shown that two pairs of triplet Higgs
are needed to provide the $ CP $ violating phase for leptogenesis.
Furthermore, the masses of triplet Higgs should be almost degenerate
in order to produce the sufficient lepton number asymmetry.
Hence, it seems that the successful leptogenesis
is obtained in a rather restricted situation
in this triplet Higgs decay scenario.

As an alternative possibility, we investigate in this article
the Affleck-Dine leptogenesis in an extension
of the minimal supersymmetric standard model
including a pair of triplet Higgs fields $ \Delta $ and $ {\bar \Delta} $.
The gauge singlet mass $ M_\Delta $ of triplet Higgs may be much larger
than the gravitino mass $ m_{3/2} \sim 10^3 {\rm GeV} $
representing the low-energy soft supersymmetry breaking terms.
There appear some new interesting features in this leptogenesis scenario.
Specifically, the lepton number asymmetry
is generated on a multi-dimensional flat manifold consisting
of the triplet Higgs $ \Delta $, $ {\bar \Delta} $
and anti-slepton $ {\tilde e}^c $.
This flat manifold is spanned by the two directions,
the one is represented by $ {\bar \Delta} \Delta $ ($ Q_L = 0 $)
and the other by $ {\bar \Delta} {\bar \Delta} {\tilde e}^c {\tilde e}^c $
($ Q_L \not= 0 $).
These directions are comparably flat
with the superpotential terms
$ ( \lambda_{L \! \! \! /} / 2M ) {\bar \Delta} {\bar \Delta} e^c e^c $
and $ ( \lambda_{\Delta} / 2M ) {\bar \Delta} \Delta {\bar \Delta} \Delta $
of the same order (though any specific relation need not be assumed
between $ \lambda_{L \! \! \! /} $ and $ \lambda_{\Delta} $).
It is the essential point that there are several potential terms
depending differently on the phases of Affleck-Dine (AD) fields
$ \Delta $, $ {\bar \Delta} $, $ {\tilde e}^c $
which are significant for the Hubble parameter
$ H \gtrsim M_\Delta \gg m_{3/2} $.
Then, as shown in detail in the text,
due to the effects of such terms the phases of AD fields
start to fluctuate soon after the inflation,
and then they are rotated at the time with $ H \sim M_\Delta $.
The lepton number asymmetry is generated through this time variation
of the phases of AD fields with the lepton number violation
provided by the superpotential term $ {\bar \Delta} {\bar \Delta} e^c e^c $.
The fraction of the resultant lepton number asymmetry
amounts to $ \epsilon_L \sim 0.1 $ for the generic model parameter values
with $ \lambda_{L \! \! \! /} / \lambda_{\Delta} \sim 0.3 - 3 $,
which is rather independent of $ M_\Delta $.
Therefore, the lepton number asymmetry for the successful baryogenesis
is indeed generated even for $ M_\Delta \gg m_{3/2} $
in the present scenario,
which can be sufficiently before the thermal effects take place.

\section*{Model}

Our leptogenesis scenario is implemented in an extension
of the minimal supersymmetric standard model
by introducing a pair of triplet Higgs superfields
with gauge anomaly cancellation.
The triplet Higgs superfields are listed as follows
with their quantum numbers of
$ {\rm SU(3)}_C \times {\rm SU(2)}_{\rm L} \times {\rm U(1)}_Y $:
%%%%%
\begin{eqnarray}
\Delta = ( \Delta^{++} , \Delta^+ , \Delta^0 )
= ({\bf 1},{\bf 3},1) , \
{\bar \Delta} = ( {\bar \Delta}^0 , {\bar \Delta}^- , {\bar \Delta}^{--} )
= ({\bf 1},{\bf 3},-1) .
\end{eqnarray}
%%%%%
The generic $ R $-parity preserving superpotential
relevant for the leptons and Higgs is given by
%%%%%
\begin{eqnarray}
W = h L H_d e^c + \mu H_u H_d
  + {\rm e}^{i \delta_{M_\Delta}} M_\Delta {\bar \Delta} \Delta
  + f_1 \Delta L L + f_2 \Delta H_d H_d + f_3 {\bar \Delta} H_u H_u ,
\label{W}
\end{eqnarray}
%%%%%
where the generation index is suppressed for simplicity.
We also assume $ R $-parity preserving non-renormalizable terms,
%%%%%
\begin{eqnarray}
W_{\rm non} = \frac{\lambda_{L \! \! \! /}}{2M}
{\bar \Delta} {\bar \Delta} e^c e^c
+ \frac{\lambda_\Delta}{2M} {\bar \Delta} \Delta {\bar \Delta} \Delta ,
\label{Wnon}
\end{eqnarray}
%%%%%
where $ M $ represents some very large mass scale
such as the grand unification or Planck scale.
For convenience of later analysis on leptogenesis,
the coupling constants $ \lambda_{L \! \! \! /} $
and $ \lambda_\Delta $ are taken to be real
by making suitable phase transformations of the relevant fields,
while the complex phase is explicitly factored out
in the $ {\bar \Delta} \Delta $ term with $ M_\Delta > 0 $.
Other non-renormalizable terms may exist as well,
though they are not relevant for the leptogenesis
investigated here.

The triplet Higgs fields are $ R $-parity even, and their lepton numbers
are assigned as
%%%%%
\begin{eqnarray}
Q_L ( \Delta ) = -2 , \ Q_L ( {\bar \Delta} ) = 2 .
\label{LDelta}
\end{eqnarray}
%%%%%
Then, the triplet Higgs mass term $ M_\Delta $
and the $ f_1 $ coupling are lepton number conserving,
while the lepton number violation is provided
by the $ f_2 $ and $ f_3 $ couplings
and also the $ \lambda_{L \! \! \! /} $ term in $ W_{\rm non} $.

It is noticed in Eq. (\ref{W})
that the triplet Higgs $\Delta$ couples to the two lepton doublets
$ LL $.  Hence, if $\Delta^0$ develops a vev (vacuum expectation value),
then the ordinary neutrinos acquire Majorana masses
through this $ f_1 $ coupling.
In fact, by checking the vanishing of $F$-terms
we find that in the supersymmetric limit
the following vev's are induced for the triplet Higgs fields
through the electroweak gauge symmetry breaking:
%%%%%
\begin{eqnarray}
\langle {\bar \Delta}^0 \rangle
= - \frac{ f_2 \langle H_d \rangle^2 }{M_\Delta}
  {\rm e}^{- i \delta_{M_\Delta}} , \
\langle {\Delta}^0 \rangle
= - \frac{ f_3 \langle H_u \rangle^2 }{M_\Delta}
  {\rm e}^{-i \delta_{M_\Delta}} ,
\end{eqnarray}
%%%%%
where $ v \equiv [ \langle H_u \rangle^2 + \langle H_d \rangle^2 ]^{1/2}
= 174 {\rm GeV} $.
The $ f_2 $ coupling combined with the $ \mu $ term
provides contributions $ \sim f_2 \mu v^2 / M_\Delta^2 $ to these vev's,
and the soft supersymmetry breaking terms also provide contributions
$ \sim m_{3/2} v^2 / M_\Delta^2 $.
(We do not consider below these sub-leading contributions for simplicity.)
It should here be noted that these vev's are induced
by the couplings explicitly violating the lepton number conservation.
Hence, the so-called triplet Majoron does not appear
from the $ \Delta $ and $ {\bar \Delta} $ fields,
which rather acquire masses $ \simeq M_\Delta $.

The mass matrix for the ordinary neutrinos is given by
%%%%%
\begin{eqnarray}
( m_{\nu} )_{ij} = f_1^{ij} f_3 \frac{\langle H_u \rangle^2}{M_\Delta} .
\end{eqnarray}
%%%%%
This neutrino mass matrix is expected to provide an eigenvalue
$ \sim 10^{-2} {\rm eV} $ and smaller ones,
in particular, in order to resolve the atmospheric neutrino anomaly.
Then, the following condition is required
for the eigenvalues $ ( f_1 )_i $ of $ f_1 $,
%%%%%
\begin{eqnarray}
( f_1 )_i f_3
\lesssim 10^{-4} \left( \frac{M_\Delta}{10^{11} \rm GeV} \right) .
\label{fcond}
\end{eqnarray}
%%%%%
This condition may be realized for reasonable values
of $ f_1 , f_3 \sim 10^{-2} $,
if the triplet Higgs fields are very heavy
with $ M_\Delta \sim 10^{11} {\rm GeV} $
\cite{triseesaw}.
It should be noted that the triplet Higgs mass
can be heavy as long as the condition
$ M_\Delta \ll H_{\rm inf} $ (the Hubble parameter during the inflation)
is satisfied for the Affleck-Dine leptogenesis with triplet Higgs.

An alternative way to realize the condition (\ref{fcond})
from the neutrino masses may be considered,
which is valid even for $ M_\Delta \sim m_{3/2} $.
That is, the lepton number violation may originate
in the ultra high energy scale $ M $
such as the grand unification or Planck scale.
Then, the effective lepton number violation will appear
as non-renormalizable terms in the electroweak scale,
being suppressed by this large scale $ M $.
Specifically, the $ f_2 $ and $ f_3 $ terms
may be replaced by the following ones,
%%%%%
\begin{eqnarray}
W_{\rm non}^\prime = f_2^\prime \frac{S \Delta H_d H_d }{M}
+ f_3^\prime \frac{S {\bar \Delta} H_u H_u}{M} .
\end{eqnarray}
%%%%%
Here the gauge singlet field $ S $ with mass $ \sim m_{3/2} $
is also assumed to be present in the electroweak scale.
This singlet field $ S $ may develop a vev
$\langle S \rangle \sim m_{3/2} \sim 10^3 {\rm GeV} $.
Then, the $ f_2 $ and $ f_3 $ couplings are induced effectively as
%%%%%
\begin{eqnarray}
f_2 = f_2^\prime \frac{\langle S \rangle }{M} , \
f_3 = f_3^\prime \frac{\langle S \rangle }{M} .
\end{eqnarray}
%%%%%
Now, if $M \sim10^{16} {\rm GeV} $ or larger,
the condition (\ref{fcond}) for the neutrino masses is satisfied
for the reasonable values of $ f_1 $ and $ f_3^\prime $
even in the case of $ M_\Delta \sim m_{3/2} \sim 10^3 {\rm GeV} $.

In any case, it is reasonably expected
that the lepton number violating couplings $ f_2 $ and $ f_3 $
are much smaller than the lepton number conserving coupling $ f_1 $.
Then, the triplet Higgs $ \Delta $ ($ {\bar \Delta} $)
decays predominantly to two anti-leptons (sleptons)
through the $ f_1 $ coupling.
Therefore, the triplet Higgs fields are considered to carry
almost the definite lepton numbers as assigned in Eq. (\ref{LDelta}).
This feature is relevant for the Affleck-Dine leptogenesis
with triplet Higgs, since after the leptogenesis the lepton number
should be conserved with sufficient accuracy.

\section*{Flat Directions}

We assume that the triplet Higgs mass is negligible
at the end of inflation,
%%%%%
\begin{eqnarray}
M_\Delta \ll H_{\rm  inf} ,
\end{eqnarray}
%%%%%
though it may be much larger than the gravitino mass,
$ M_\Delta \gg m_{3/2} $.
Then, the triplet Higgs fields are allowed to enter into
certain flat directions until the Hubble parameter $ H $
decreases to $M_\Delta$.
(This situation is similar to the $ \mu $ term
in the case of $ {\tilde L} H_u $ flat direction.)
The $F$-terms are actually given (except for the contributions
of $ W_{\rm non} $) as
%%%%%
\begin{eqnarray}
 F_L        &=& h H_d e^c + 2 f_1 \Delta L ,
\nonumber \\
 F_{e^c}    &=& h L H_d ,
\nonumber \\
 F_{H_u}    &=& \mu H_d  + 2 f_3 {\bar \Delta} H_u ,
\nonumber \\
 F_{H_d}    &=& h L e^c + \mu H_u + 2 f_2 \Delta H_d ,
\nonumber \\
 F_{\Delta} &=& {\rm e}^{i \delta_{M_\Delta}} M_\Delta {\bar \Delta}
             + f_1 L L + f_2 H_d H_d ,
\nonumber \\
 F_{{\bar \Delta}} &=& {\rm e}^{i \delta_{M_\Delta}} M_\Delta \Delta
                    + f_3 H_u H_u .
\end{eqnarray}
%%%%%
The flat directions involving the triplet Higgs fields
are specified by the $ D $-flat condition
%%%%%
\begin{eqnarray}
| \Delta^+ |^2 - | {\bar \Delta}^- |^2 + | {\tilde e}^c |^2 = 0 ,
\label{flat}
\end{eqnarray}
%%%%%
and the other fields are all vanishing
for the $ F $-flat conditions up to the contributions
of $ M_\Delta $.
(We consider in the following the one generation of $ {\tilde e}^c $
for simplicity.)
The corresponding gauge singlet combinations
are $ {\bar \Delta} \Delta $ ($ Q_L = 0 $)
and $ {\bar \Delta} {\bar \Delta} {\tilde e}^c {\tilde e}^c $
($ Q_L = 2 $), as given in Eq. (\ref{Wnon}).
It should here be noted that these directions may be comparably flat,
if the relevant non-renormalizable superpotential terms
are of the same order,
specifically $ \lambda_{L \! \! \! /} / \lambda_\Delta \sim 0.3 - 3 $,
as seen explicitly later in the numerical analysis.
Then, in contrast to the original Affleck-Dine scenario,
the evolution of the scalar fields takes place
on the complex two-dimensional flat manifold spanned by these directions.

\section*{Affleck-Dine Leptogenesis}

We now examine the lepton number asymmetry
generated by the Affleck-Dine mechanism with triplet Higgs
in the present model.
The relevant flat manifold is specified in Eq. (\ref{flat}).
The scalar potential for the AD fields
$ \Delta^+ $, $ {\bar \Delta}^- $, $ {\tilde e}^c $ is given by
%%%%%
\begin{eqnarray}
V
&=& ( C_1 m_{3/2}^2 - c_1 H^2 ) |\Delta|^2
 + ( C_2 m_{3/2}^2 - c_2 H^2 ) |{\bar \Delta}|^2
\nonumber \\
&+& ( C_3 m_{3/2}^2 - c_3 H^2 ) |{\tilde e}^c|^2
\nonumber \\
&+& \left| {\rm e}^{i \delta_{M_\Delta}} M_\Delta \Delta
 + \frac{\lambda_{L \! \! \! /}}{M} {\bar \Delta} {\tilde e}^c {\tilde e}^c
 + \frac{\lambda_\Delta}{M} {\bar \Delta} \Delta \Delta \right|^2
\nonumber \\
&+& \left| {\rm e}^{i \delta_{M_\Delta}} M_\Delta {\bar \Delta}
 + \frac{\lambda_\Delta}{M} {\bar \Delta} {\bar \Delta} \Delta \right|^2
 + \left| \frac{\lambda_{L \! \! \! /}}{M}
         {\bar \Delta} {\bar \Delta} {\tilde e}^c \right|^2
\nonumber \\
&+& \left[ ( b_\Delta H +  B_\Delta m_{3/2} )
    M_\Delta {\bar \Delta} \Delta
 + {\rm h.c.} \right]
\nonumber \\
&+& \left[ \frac{1}{2M}
   ( a_{L \! \! \! /} H + A_{L \! \! \! /} m_{3/2} )
    \lambda_{L \! \! \! /}
    {\bar \Delta} {\bar \Delta} {\tilde e}^c {\tilde e}^c
 + {\rm h.c.} \right]
\nonumber \\
&+& \left[ \frac{1}{2M}
   ( a_\Delta H + A_\Delta m_{3/2} )
   \lambda_\Delta {\bar \Delta} \Delta {\bar \Delta} \Delta
  + {\rm h.c.} \right]
\nonumber \\
&+& g^{\prime 2 } (|\Delta|^2 - |{\bar \Delta}|^2 + |{\tilde e}^c|^2)^2
\label{V}
\end{eqnarray}
%%%%%
($ \Delta \equiv \Delta^+ $ and $ {\bar \Delta} \equiv {\bar \Delta}^- $
henceforth for simplicity of notation).
The last term with $ {\rm U(1)}_Y $ gauge coupling $ g^\prime $
is included to realize the $ D $-flat condition (\ref{flat})
for the large enough $ |\Delta| $, $ |{\bar \Delta}| $, $ |{\tilde e}^c| $.
The nonzero energy density in the early universe provides
the soft supersymmetry breaking terms with the Hubble parameter $ H $
in addition to the low-energy ones with the gravitino mass $ m_{3/2} $
\cite{DRT}.  In the following, the evolution of the AD fields is described
in the respective epochs for evaluating the lepton number asymmetry.
It will really be confirmed later by solving the equations of motion
numerically with the initial values specified by the potential minimum
in the inflation epoch.

\subsection*{(i) $ t \ll M_\Delta^{-1} $}

During the inflation the Hubble parameter takes almost
a constant value $ H_{\rm inf} $,
and the AD fields quickly settle into one of the minima
of the scalar potential $ V $ with $ H = H_{\rm inf} $:
%%%%%
\begin{equation}
\phi_a^{(0)} = {\rm e}^{i \theta_a^{(0)}} r_a^{(0)}
\sqrt{{H_{\rm inf} (M/\lambda)}} ,
\label{phi0}
\end{equation}
%%%%%
where $ \phi_a = \Delta , {\bar \Delta} , {\tilde e}^c $,
and $ \lambda $ represents the mean value
of $ \lambda_{L \! \! \! /} $ and $ \lambda_\Delta $.
These minima are determined depending on the values of the parameters
in the scalar potential.
Specifically, the $ \lambda_{L \! \! \! /} $-$ \lambda_\Delta $ cross term,
$ a_{L \! \! \! /} $ term and $ a_\Delta $ term,
which are significant for $ H \gg M_\Delta , m_{3/2} $,
have different dependences on the phases $ \theta_a $ of AD fields.
Then, some valleys are formed on the flat manifold
so as to minimize the sum of these three terms
depending generally on $ | \phi_a | $,
and the potential minima are located along them.
Actually, we find the minima with
%%%%%
\begin{equation}
r_a^{(0)} \sim 0.1 - 1
\leftarrow \lambda_{L \! \! \! /} \sim \lambda_\Delta , \
c_a \sim 1 .
\end{equation}
%%%%%
(It should be mentioned for completeness
that if $ \lambda_{L \! \! \! /} $ and $ \lambda_\Delta $
are rather different from each other, the minimum is formed
along one of the flat directions
with $ r_{\Delta}^{(0)} = 0 $ or $ r_{{\tilde e}^c}^{(0)} = 0 $.
We do not consider such cases here.)
As for the initial phases,
up to the physically irrelevant arbitrary phase
of $ {\rm U(1)}_{\rm em} $ gauge transformation they are specified
in terms of the $ {\rm U(1)}_{\rm em} $ invariant combinations as
%%%%%
\begin{equation}
\theta_\Delta^{(0)} + \theta_{\bar \Delta}^{(0)} , \
\theta_{{\tilde e}^c}^{(0)} + \theta_{\bar \Delta}^{(0)} .
\end{equation}
%%%%%
The dependence of $ \theta_a^{(0)} $ on the parameters
$ c_a, \lambda_{L \! \! \! /} , \lambda_\Delta ,
a_{L \! \! \! /} , a_\Delta $
is really complicated in contrast to the original one-dimensional
Affleck-Dine scenario, unless the fine-tuning,
$ \arg ( a_{L \! \! \! /} ) - \arg ( a_\Delta ) = \pi \ {\rm mod} \ 2 \pi $,
is made so as to align simultaneously
the three phase-dependent potential terms.

After the inflation the inflaton oscillates coherently,
and it dominates the energy density of the universe.
In this epoch of $ t \ll M_\Delta^{-1} < m_{3/2}^{-1} $
($ H \gg M_\Delta > m_{3/2} $), the AD fields are moving toward the origin
with the initial conditions at $ t = t_0 \sim H_{\rm inf}^{-1}$
just after the inflation,
%%%%%
\begin{equation}
\phi_a (t_0) = \phi_a^{(0)} , \ {\dot \phi}_a (t_0) = 0 .
\label{phit0}
\end{equation}
%%%%%
The evolution of the AD fields are governed by the equations of motion,
%%%%%
\begin{equation}
\ddot{\phi}_a + 3 H \dot{\phi}_a
+ \frac{\partial V}{\partial \phi_a^*} = 0 .
\label{eqofm}
\end{equation}
%%%%%
The Hubble parameter varies in time as $ H = (2/3) t^{-1} $
in the matter-dominated universe.
The AD fields may be represented suitably
in terms of the dimensionless fields $ \chi_a $
\cite{DRT} as
%%%%%
\begin{eqnarray}
\phi_a = \chi_a \sqrt{{H (M/\lambda)}}
\equiv {\rm e}^{i \theta_a} r_a \sqrt{{H (M/\lambda)}} .
\label{phi}
\end{eqnarray}
%%%%%
Then, the equations of motion (\ref {eqofm}) are rewritten
with $ z = \ln ( t / t_0 ) $ as
%%%%%
\begin{equation}
\frac{d^2 \chi_a}{dz^2} + \frac{\partial U}{\partial \chi_a^*} = 0 ,
\label{eqofmchi}
\end{equation}
%%%%%
and the initial conditions from Eq. (\ref{phit0}) are given as
%%%%%
\begin{equation}
\chi_a (0) = {\rm e}^{i \theta_a^{(0)}} r_a^{(0)} , \
\frac{d \chi_a}{dz} (0) = \frac{1}{2} \chi_a (0) .
\label{chi0}
\end{equation}
%%%%%
The dimensionless effective potential $ U $ is given by
%%%%%
\begin{equation}
U ( \chi_a , M_\Delta / H , m_{3/2} / H )
= \frac{4}{9 H^3 ( M / \lambda)} V ( \phi_a , H , M_\Delta , m_{3/2} )
- \frac{1}{4} | \chi_a |^2 .
\end{equation}
%%%%%
The second term is due to the time variation of the factor
$ \sqrt{{H (M/\lambda)}} $ in Eq. (\ref{phi}),
which apparently provides the change of the mass terms in $ U $,
%%%%%
\begin{equation}
c_a \rightarrow c_a + \frac{9}{16} .
\label{dca}
\end{equation}
%%%%%
It should be noticed in Eq. (\ref{eqofmchi})
that the first-order $ z $-derivative is absent
due to the parameterization of $ \phi_a \propto H^{1/2} \propto t^{-1/2} $
in Eq. (\ref{phi}).
In this epoch with $ H \gg M_\Delta , m_{3/2} $, the effective potential
$ U $ is almost independent of the mass parameters $ M_\Delta , m_{3/2} $:
%%%%%
\begin{equation}
U ( \chi_a , M_\Delta / H , m_{3/2} / H )
= U_1 ( \chi_a ) + O ( M_\Delta / H , m_{3/2} / H ) .
\end{equation}
%%%%%

The motion of the phases $ \theta_a $ of AD fields is described
in this epoch as follows.  The initial conditions at $ t = t_0 $ ($ z = 0 $)
are given from Eq. (\ref{chi0}) as
%%%%%
\begin{equation}
\theta_a (0) = \theta_a^{(0)} , \ \frac{d \theta_a}{dz} (0) = 0 .
\end{equation}
%%%%%
On the other hand, the asymptotic trajectory of the AD fields
is found by the conditions $ \partial U_1 / \partial \chi_a^* = 0 $
in this epoch with $ H \gg M_\Delta , m_{3/2} $ as
%%%%%
\begin{equation}
\theta_a = \theta_a^{(1)} .
\end{equation}
%%%%%
It is remarkable for the multi-dimensional motion
of the AD fields with $ \lambda_{L \! \! \! /} \sim \lambda_\Delta $
that the direction of this trajectory is somewhat different
from the initial direction, i.e.,
%%%%%
\begin{equation}
\theta_a^{(1)} \not= \theta_a^{(0)} .
\end{equation}
%%%%%
This is because the apparent change of the mass terms in Eq. (\ref{dca})
due to the redshift induces the new balance
among the $ \lambda_{L \! \! \! /} $-$ \lambda_\Delta $ cross term,
$ a_{L \! \! \! /} $ term and $ a_\Delta $ term in $ U_1 ( \chi_a ) $,
which have different dependences on $ \theta_a $.
(If the fine-tuning is made as
$ \arg ( a_{L \! \! \! /} ) - \arg ( a_\Delta ) = \pi \ {\rm mod} \ 2 \pi $,
the initial balance is maintained independently of $ | \chi_a | $
so as to realize $ \theta_a^{(0)} = \theta_a^{(1)} $.)
Without the $ d \chi_a / d z $ (friction) term in Eq. (\ref{eqofmchi}),
the phases of AD fields $ \theta_a $ slowly fluctuate
around $ \theta_a^{(1)} $ starting from $ \theta_a^{(0)} $
as a function of $ z = \ln ( t / t_0 ) $
in the epoch $ H_{\rm inf}^{-1} \sim t_0 \leq t < M_\Delta^{-1} $.
That is, in the motion on the multi-dimensional flat manifold
the AD fields no longer track exactly behind the decreasing
instantaneous minimum of scalar potential $ V $.
This is a salient contrast to the usual Affleck-Dine mechanism
on the one-dimensional flat direction,
where the phase of one AD field is kept constant
until it begins to oscillate by the low-energy supersymmetry breaking
mass term or the thermal mass term.
In this way, even in this very early epoch the lepton number asymmetry
really appears due to this phase fluctuation of the AD fields
on the multi-dimensional flat manifold.

The lepton number asymmetry is evaluated
by combining the contributions of the AD fields as
%%%%%
\begin{eqnarray}
n_L = 2 \Delta n_{\bar \Delta} - 2 \Delta n_{\Delta}
      - \Delta n_{{\tilde e}^c} ,
\end{eqnarray}
%%%%%
where the particle number asymmetry
($ {\mbox{particl number}} - {\mbox{anti-particle number}} $) is calculated
with the homogeneous coherent scalar field $ \phi_a (t) $ by
%%%%%
\begin{eqnarray}
\Delta n_a \equiv n_a - {\bar n}_a
= i ( \phi_a^* {\dot \phi}_a - {\dot \phi}_a^* \phi_a ) .
\end{eqnarray}
%%%%%
The resultant lepton number asymmetry is given as
%%%%%
\begin{eqnarray}
n_L (t) = \epsilon_L (t) (3/2) H^2 (M/\lambda)
\label{nL}
\end{eqnarray}
%%%%%
in terms of the parameter $ \epsilon_L (t) $ representing
the fraction of the lepton number asymmetry
%%%%%
\begin{equation}
\epsilon_L (t) = \sum_a Q_L (a) \epsilon_a (t)
\label{epL}
\end{equation}
%%%%%
with the respective fractions of particle number asymmetries
%%%%%
\begin{equation}
\epsilon_a (t)
= i \left( \chi_a^* \frac{d \chi_a}{dz}
           - \frac{d \chi_a^*}{dz} \chi_a \right)
= - 2 r_a^2 \frac{d \theta_a}{dz} .
\label{epa}
\end{equation}
%%%%%
Since the phases of AD fields are fluctuating in this early epoch,
as mentioned so far, the lepton number asymmetry
is oscillating in time as $ | \epsilon_L (t) | \sim | d \theta_a / dz |
\lesssim | \theta_a^{(0)} - \theta_a^{(1)} | \sim 0.01 - 0.1 $
($ r_a \sim 0.1 - 1 $) numerically for the reasonable parameter values.

\subsection*{(ii) $ t \sim M_\Delta^{-1} $}

The Hubble parameter eventually decreases after the inflation,
and when it becomes as
%%%%%
\begin{eqnarray}
H \sim M_\Delta \ ( t \sim M_\Delta^{-1} ) ,
\end{eqnarray}
%%%%%
the AD fields start to oscillate due to the triplet Higgs mass terms
$ M_\Delta^2 ( | \Delta |^2 + | {\bar \Delta} |^2 ) $.
The significant torque is also applied to the AD fields
by the phase-dependent potential terms which are provided
with the superpotential terms $ {\bar \Delta} \Delta $,
$ {\bar \Delta} {\bar \Delta} e^c e^c $,
$ {\bar \Delta} \Delta {\bar \Delta} \Delta $.
Then, in this epoch the AD fields are rotating around the origin,
and the lepton number asymmetry soon approaches
certain nonzero value as
%%%%%
\begin{eqnarray}
\epsilon_L (t) \approx \epsilon_L \ ( t \gg M_\Delta^{-1} ) .
\label{eL}
\end{eqnarray}
%%%%%
This final lepton number asymmetry is calculated
to be $ \epsilon_L \sim 0.1 $
for the generic choice of the model parameter values
with $ \lambda_{L \! \! \! /} / \lambda_{\Delta} \sim 0.3 - 3 $,
as shown later in the numerical analysis.
It should here be noted that
once the AD fields are rotated rapidly with frequency $ \sim M_\Delta $,
the low-energy soft supersymmetry breaking terms
have little effects on the leptogenesis for $ M_\Delta \gg m_{3/2} $.

\subsection*{(iii) $ t \gg M_\Delta^{-1} $}

The coherent oscillation of the inflaton field dominates
the energy density of the universe until the decay of the inflatons
is completed at the time $ t_R $ ($ \gg M_\Delta^{-1} $).
Then, the universe is reheated to the temperature $ T_R $.
Until this time, the lepton number asymmetry is redshifted as matter,
which is given for $ H = H_R $ with Eqs. (\ref{nL}) and (\ref{eL}) as
%%%%%
\begin{eqnarray}
n_L(t_R) = \epsilon_L (3/2) H_R^2 (M/\lambda) .
\end{eqnarray}
%%%%%
Then, the lepton-to-entropy ratio after the reheating
is estimated with $ s \sim 3 H_R^2 M_{\rm P}^2 / T_R $ as
%%%%%
\begin{eqnarray}
\frac{n_L}{s}
& \sim & \epsilon_L \frac{(M/\lambda) T_R}{2 M_{\rm P}^2}
\nonumber \\
& \sim & 10^{-10}
    \left( \frac{\epsilon_L}{0.1} \right)
    \left( \frac{10^{-2}}{\lambda} \right)
    \left( \frac{M}{10^{18} \rm GeV} \right)
    \left( \frac{T_R}{10^8 \rm GeV} \right) ,
\label{kinji}
\end{eqnarray}
%%%%%
where $ M_{\rm P} = m_{\rm P} / {\sqrt{8 \pi}}
= 2.4 \times 10^{18} {\rm GeV} $ is the reduced Planck mass.
This lepton number asymmetry is converted partially
to the baryon number asymmetry through the electroweak anomalous effect.
The chemical equilibrium between leptons and baryons leads
the ratio $ n_B \simeq - 0.35 n_L $
(without any preexisting baryon number asymmetry)
\cite{hatu}.
Therefore, the sufficient baryon-to-entropy ratio is provided
as required from the nucleosynthesis
\cite{PDG},
%%%%%
\begin{eqnarray}
\eta = ( 1.2 - 5.7 ) \times 10^{-10} .
\end{eqnarray}
%%%%%
One may take seriously the non-thermal gravitino production.
Then, the reheating temperature may be significantly lower
than $10^8 $ GeV
\cite{ntgra}.
Even in this case, if $\lambda$ is small enough, 
the sufficient baryon number asymmetry can be generated.

\subsection*{Numerical analysis}

We have made numerical calculations
to confirm the generation of lepton number asymmetry
in the present model with triplet Higgs.
The values of the model parameters are taken in some reasonable ranges as
%%%%%
\begin{eqnarray}
&& M = 10^{18} {\rm GeV} , \ M_\Delta = m_{3/2} - 0.1 H_{\rm inf} , \
m_{3/2} = 10^3 {\rm GeV} ,
\nonumber \\
&& \lambda_{L \! \! \! /} , \lambda_\Delta = 0.3 \lambda - 3 \lambda , \
\lambda = 10^{-2} ,
\nonumber \\
&& c_a , C_a = 0.5 - 2 ,
\nonumber \\
&& | a_{L \! \! \! /} | , | a_\Delta | , | b_\Delta | ,
|A_{L \! \! \! /}| , |A_\Delta| , |B_\Delta| = 0.5 - 2 ,
\label{range}
\end{eqnarray}
%%%%%
and $ [ 0 , 2 \pi ] $ for the phases of coupling parameters.
A typical example for the evolution of the AD fields is presented
in the following by taking the parameter values rather arbitrarily
in the above ranges as
%%%%%
\begin{eqnarray}
&& M_\Delta = 10^{-4} H_{\rm inf}
= 10^9 {\rm GeV} ( H_{\rm inf} / 10^{13} {\rm GeV} ) ,
\nonumber \\
&& \lambda_{L \! \! \! /} = \lambda , \
\lambda_\Delta = 2 \lambda ,
\nonumber \\
&& c_{\bar{\Delta}} = 1 , \ c_{\Delta} = 0.5 , \ c_{{\tilde e}^c} = 1.5 , \
\nonumber \\
&& | a_{L \! \! \! /} | = 1 \ , | a_\Delta | = 0.5 , \ | b_\Delta | = 1 , \
\nonumber \\
&& \arg ( a_{L \! \! \! /} ) = - \pi / 6 , \ \arg (a_\Delta) = - 2 \pi / 3 ,
\nonumber \\
&& \arg ( b_\Delta ) =  5 \pi / 6 , \ \delta_{M_\Delta} = \pi / 3 .
\label{example}
\end{eqnarray}
%%%%%
(The effects of the low-energy soft supersymmetry breaking terms
are in fact negligible for $ M_\Delta \gg m_{3/2} $.)
The initial values of the AD fields at $ t = t_0 $
($ H = H_{\rm inf} \gg M_\Delta , m_{3/2} $)
in Eq. (\ref{phi0}) are determined with these parameter values as
%%%%%
\begin{eqnarray}
&& r_{\bar{\Delta}}^{(0)} = 0.752 , \ r_\Delta^{(0)} = 0.144 , \
r_{\tilde{e}^c}^{(0)} = 0.738 ,
\nonumber \\
&& \theta_{\bar{\Delta}}^{(0)} = 0 , \ \theta_\Delta^{(0)} = - 3.044 , \
\theta_{\tilde{e}^c}^{(0)} = - 1.327 ,
\label{phi0a}
\end{eqnarray}
%%%%%
where $ \theta_{\bar{\Delta}}^{(0)} = 0 $ is chosen
by the $ {\rm U(1)}_{\rm em} $ gauge transformation.
Then, the asymptotic trajectory of the AD fields
in the epoch $ t_0 < t \ll M_\Delta^{-1} < m_{3/2}^{-1} $ is determined
by the conditions $ \partial U_1 / \partial \chi_a^* = 0 $ as
%%%%%
\begin{eqnarray}
&& r_{\bar{\Delta}}^{(1)} = 0.812 , \ r_\Delta^{(1)} = 0.209 , \
r_{\tilde{e}^c}^{(1)} = 0.786 ,
\nonumber \\
&& \theta_{\bar{\Delta}}^{(1)} = 0.006 , \ \theta_\Delta^{(1)} = - 3.048 , \
\theta_{\tilde{e}^c}^{(1)} = - 1.348 .
\label{phi1a}
\end{eqnarray}
%%%%%
It is really observed that the asymptotic phases $ \theta_a^{(1)} $ 
are in general slightly different from the initial phases $ \theta_a^{(0)} $
in the multi-dimensional motion of the AD fields
due to the apparent change of the mass terms in Eq. (\ref{dca}).
(It is also checked that if the fine-tuning,
$ \arg ( a_{L \! \! \! /} ) - \arg ( a_\Delta ) = \pi \ {\rm mod} \ 2 \pi $,
is made, the alignment $ \theta_a^{(1)} = \theta_a^{(0)} $ is realized,
as expected.)

We have solved numerically the equations of motion (\ref{eqofm})
for the AD fields from $ t = t_0 $ ($ H = H_{\rm inf} $)
to $ t \sim 100 M_\Delta^{-1} $ with the initial conditions
at $ t = t_0 $ in Eq. (\ref{phit0}).
(In practice, we have solved Eq. (\ref{eqofmchi}) for $ \chi_a $
with Eq. (\ref{chi0}) as functions of $ z = \ln (t/t_0) $
since the time interval ranges over several orders.
The $ D $-flat condition (\ref{flat}) is checked to be hold.)
The result is depicted in Fig. \ref{tra}
in terms of the dimensionless fields $ \chi_a $
for the case of Eq. (\ref{example})
with $ \lambda_{L \! \! \! /} / \lambda_\Delta = 0.5 $
and $ M_\Delta = 10^{-4} H_{\rm inf} $ ($ \gg m_{3/2} $).
The dots represent the times of
$ t / t_0 = 1 , 10 , 10^2 , 10^3 , 10^4 , 10^5 $.
The AD fields really exhibit the behavior
as described in the preceding section.
Their phases $ \theta_a $ and magnitudes
$ r_a = | \chi_a | $ normalized in Eq. (\ref{phi})
fluctuate gradually around the asymptotic values
in Eq. (\ref{phi1a}) for $ t_0 < t < 10^4 t_0 $.
Then, around $ t \sim 10^4 t_0 \sim M_\Delta^{-1} $
they begin to rotate around the origin,
which appears to be somewhat complicated in the multi-dimensional motion.
This motion of the AD fields for $ t \gtrsim M_\Delta^{-1} $
is driven mainly by the potential terms provided with the superpotential term
$ M_\Delta {\bar \Delta} \Delta $ of triplet Higgs mass.

\begin{figure}[ht]
\begin{center}
\includegraphics*[0cm,0cm][24.0cm,15.5cm]{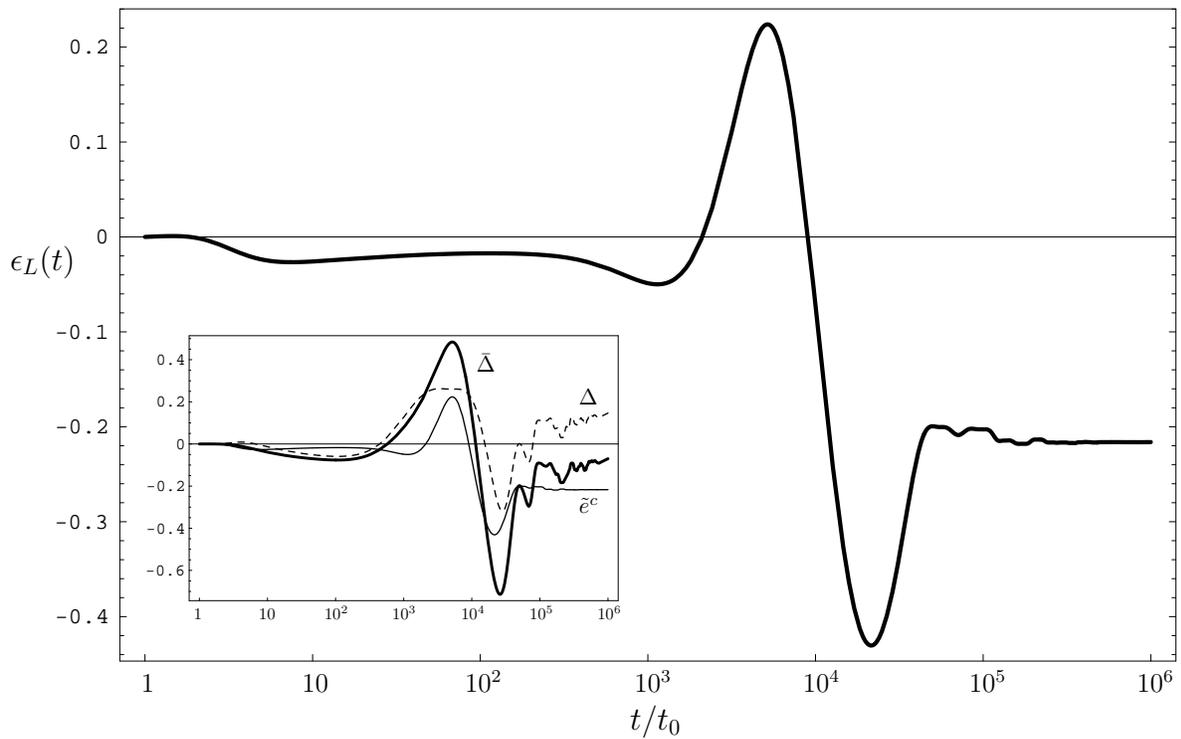}
\vspace*{0.5cm}
\caption{The fraction of lepton number asymmetry $ \epsilon_L (t) $
is shown together with the respective particle number asymmetries
$ \epsilon_a (t) $
for the case with $ \lambda_{L \! \! \! /} / \lambda_\Delta = 0.5 $
and $ M_\Delta = 10^{-4} H_{\rm inf} \gg m_{3/2} $.
}
\label{epsilonL}
\end{center}
\end{figure}

The fraction of lepton number asymmetry $ \epsilon_L (t) $
in Eq. (\ref{epL})
calculated from the time evolution of the AD fields
is shown in Fig. \ref{epsilonL}
together with the respective particle number asymmetries $ \epsilon_a (t) $
in Eq. (\ref{epa}).
(It is checked that the electric charge conservation is hold
with $ \epsilon_\Delta (t) - \epsilon_{\bar \Delta} (t)
+ \epsilon_{{\tilde e}^c} (t) = 0 $.)
The lepton number asymmetry is really oscillating slowly
in the time range $ t_0 < t < 10^4 t_0 $,
which is due to the motion of the AD fields fluctuating
around the asymptotic trajectory.
Then, it changes to approach a nonzero value $ \epsilon_L \sim 0.1 $
for $ t \gtrsim 10^4 t_0 \sim M_\Delta^{-1} $.

We have obtained similar numerical results
in most cases by taking randomly about one hundred samples of the parameters
in the ranges of Eq. (\ref{range}).
Then, we have confirmed that
the minimum during the inflation for the initial values of the AD fields
can really be formed on the flat manifold with
%%%%%
\begin{equation}
r_{\bar{\Delta}}^{(0)} , r_\Delta^{(0)} , r_{\tilde{e}^c}^{(0)}
\sim 0.1 - 1 .
\end{equation}
%%%%%
It is essential for admitting this sort of multi-dimential motion
of the AD fields that the relevant non-renormalizable superpotential
terms are comparable, specifically in the present model
%%%%%
\begin{equation}
0.3 \lesssim \lambda_{L \! \! \! /} / \lambda_\Delta \lesssim 3
\end{equation}
%%%%%
with rather arbitrary values $ 0 < c_a \lesssim 1 $
for the Hubble induced mass terms.
This desired range of $ \lambda_{L \! \! \! /} / \lambda_\Delta $
has actually a reasonable size, in contrast to naive expectation,
due to the effects of the phase dependent potential terms.
If the difference between these couplings is larger than this range,
we have found that the AD fields evolve along one of the flat directions
as in the usual Affleck-Dine scenario.

As confirmed by these numerical calculations,
it is quite an interesting feature of the present Affleck-Dine leptogenesis
with triplet Higgs that the significant lepton number asymmetry is generated
in the early epoch $ t \sim M_\Delta^{-1} $
through the multi-dimensional motion of the AD fields.
Then, for $ M_\Delta \gg m_{3/2} $
the low-energy supersymmetry breaking terms with $ m_{3/2} $
have little effect on the leptogenesis .
This is because several scalar potential terms are provided
for $ H \sim M_\Delta $ with the superpotential mass term of triplet Higgs,
which have different dependences on the phases of AD fields.
The phase rotation of the AD fields for $ t \gtrsim M_\Delta^{-1} $
is indeed driven by such terms with $ M_\Delta $
rather than the low-energy soft supersymmetry breaking terms
with $ m_{3/2} $.

\section*{Thermal Effects}

We now consider the thermal effects,
which might be suspected to suppress the generation of asymmetry
\cite{hay,ACE}.

The AD fields acquire the thermal masses from the coupling
to the dilute plasma with temperature
%%%%%
\begin{equation}
T_{\rm p} \sim (T_R^2 H M_{\rm P} )^{1/4} .
\end{equation}
%%%%%
One can roughly estimate the Hubble parameter $ H_{\rm th} $
at the time when the thermal mass terms begin to dominate
over the Hubble induced mass terms:
%%%%%
\begin{eqnarray}
H_{\rm th} \sim \min \left[ \frac{T_R^2 M_{\rm P}}{y^4 ( M / \lambda )^2} ,
( y^4 T_R^2 M_{\rm P} )^{1/3} \right] ,
\end{eqnarray}
%%%%%
where $ y $ represents the couplings of the relevant fields
with the AD fields.  It takes the maximal value as
%%%%%
\begin{equation}
H_{\rm th}^{\rm max} \sim 10^7 {\rm GeV}
\left( \frac{T_R}{10^8 {\rm GeV}} \right)
\left( \frac{M / \lambda}{10^{20} {\rm GeV}} \right)^{-1/2}
\end{equation}
%%%%%
with certain value of the relevant coupling
%%%%%
\begin{equation}
y \sim 10^{-4} \left( \frac{T_R}{10^8 {\rm GeV}} \right)^{1/4}
\left( \frac{M / \lambda}{10^{20} {\rm GeV}} \right)^{-3/8} .
\label{ybar}
\end{equation}
%%%%%

We naturally consider the case of $ M_{\Delta} > H_{\rm th}^{\rm max} $
since the very large triplet Higgs mass may be desired
for the small neutrino masses, as seen in Eq. (\ref{fcond}).
Then, the triplet Higgs mass terms dominate over the thermal mass terms.
The thermal-log term
$ - a \alpha_2^2 T_p^4 \ln ( | \phi_a |^2 / T_p^2 ) $ ($ a \sim 3 C_2 = 6 $)
for the unbroken $ {\rm U(1)}_{I_3} $ in the $ {\rm SU(2)}_{\rm L} $
may act as negative mass squired term, since the gauge bosons of
$ {\rm SU(2)}_{\rm L} \times {\rm U(1)}_Y / {\rm U(1)}_{I_3} $
decouple by aquiring the large masses from the AD fields.
However, if $ M_\Delta > H_{\rm th}^{\rm max} $,
the coherent AD fields have much more energy density
than the dilute plasma, i.e.,
$ M_\Delta^2 ( | \Delta |^2 + | {\bar \Delta} |^2 )
\sim M_\Delta^2 H ( M / \lambda ) > T_{\rm p}^4 $.
Then, the thermal-log term is also much smaller
than the triplet Higgs mass terms,
and the evaporation of the AD fields is negligible energetically.
Therefore, in this preferable case of $ M_\Delta > H_{\rm th}^{\rm max} $
the leptogenesis is certainly completed
in the early epoch $ \sim M_\Delta^{-1} $
before the thermal effects become significant.

If the triplet Higgs mass is rather small
as $ M_\Delta < a^{1/2} \alpha_2 H_{\rm th}^{\rm max}
\sim 0.1 H_{\rm th}^{\rm max} $
(though not so plausible for the small neutrino masses),
the situation of leptogenesis may appear to be rather different.
In this case, the negative thermal-log term dominates
over the triplet Higgs mass terms.
On the other hand, the thermal mass terms may dominate
over the thermal-log term
if the relevant couplings satisfy the condition
$ y^2 T_{\rm p}^2 | \phi_a |^2 \sim y^2 T_{\rm p}^2 H ( M / \lambda )
> a \alpha_2^2 T_p^4 $ with $ y | \phi_a | < T_{\rm p} $.
The tyical coupling value is $ y \sim 10^{-4} $
for $ H \sim 10^6 {\rm GeV} $,  $ T_R \sim 10^8 {\rm GeV} $,
$ ( M / \lambda ) \sim 10^{20} {\rm GeV} $.
It is possible that some of the couplings
$ h , f_1 , f_2 , f_3 $ in Eq. (\ref{W}) satisfy this condition for $ y $.
In this situation, the thermal mass terms can drive the rotation
of the AD fields.
The lepton number asymmetry $ \epsilon_L (t) $ is already varying slowly
for $ t > t_0 $ soon after the inflation, as seen in Fig. \ref{epsilonL},
through the fluctuation of the AD fields
due to the several phase-dependent potential terms.
Then, if the evaporation is not significant until the AD fields
are rotated some times by the thermal mass terms, as considered in
\cite{ACE},
the lepton number asymmetry is fixed to certain nonzero value
$ \epsilon_L \sim 0.01 - 0.1 $.

To summarize, whether the thermal effects act or not
depending on $ M_\Delta $ versus $ H_{\rm th}^{\rm max} $
(while the case of $ M_\Delta > H_{\rm th}^{\rm max} $
is primarily concerned here for the small neutrino masses),
the significant lepton number asymmetry $ \epsilon_L \sim 0.01 - 0.1 $
is generally produced on the multi-dimensional flat manifold
in the present model with triplet Higgs.
This lepton number asymmetry is actually rather independent
of the value of $ M_\Delta $.

\section*{Conclusion}

We have examined the Affleck-Dine leptogenesis
in the extension of the minimal supersymmetric standard model
including a pair of triplet Higgs fields $ \Delta $ and $ {\bar \Delta} $
with mass $ M_\Delta $.
The lepton number asymmetry is generated on the multi-dimensional
flat manifold consisting of $ \Delta $, $ {\bar \Delta} $, $ {\tilde e}^c $.
It is the essential point that several phase-dependent potential terms
are provided with the superpotential terms
$ M_\Delta {\bar \Delta} \Delta $,
$ ( \lambda_{L \! \! \! /} / 2M ) {\bar \Delta} {\bar \Delta} e^c e^c $,
$ ( \lambda_{\Delta} / 2M ) {\bar \Delta} \Delta {\bar \Delta} \Delta $
($ \lambda_{L \! \! \! /} / \lambda_\Delta \sim 0.3 - 3 $),
which are significant for $ H \sim H_{\rm inf} - M_\Delta $.
Then, soon after the inflation the lepton number asymmetry appears
since the phases of AD fields fluctuate
by the effects of such potential terms.
It is slowly oscillating for a certain while,
and then the leptogenesis is completed
at the early time $ \sim M_\Delta^{-1} $,
when the AD fields begin to rotate around the origin
due to the potential terms with triplet Higgs mass.
The fraction of the resultant lepton number asymmetry
amounts in general to $ \epsilon_L \sim 0.1 $
rather independently of $ M_\Delta $.
Hence, in contrast to the usual Affleck-Dine scenario,
the low-energy soft supersymmetry breaking terms
have little effect on the leptogenesis for $ M_\Delta \gg m_{3/2} $.
The role of the thermal effects is also different in the present scenario.
The case of large triplet Higgs mass
with $ M_\Delta > H_{\rm th}^{\rm max} $
is primarily considered for the small neutrino masses.
Then, the leptogenesis is completed at the early time $ \sim M_\Delta^{-1} $
before the thermal effects become significant.
On the other hand, even if $ M_\Delta $ is rather small,
the time varying lepton number asymmetry after the inflation
is fixed to certain value $ \epsilon_L \sim 0.1 - 0.01 $
by the rotation of the AD fields due to the thermal mass terms,
which may dominate over the negative thermal-log term
with suitable values of the relevant couplings.

\newpage

\begin{figure}[ht]
\begin{center}
\includegraphics*[-3cm,0cm][10.0cm,20cm]{trajectory2.ps}
\vspace*{0.5cm}
\caption{The motions of the AD fields,
the real part (horizontal axis) and imaginary part (vertical axis),
are depicted in terms of the dimensionless fields $ \chi_a $
for the case with $ \lambda_{L \! \! \! /} / \lambda_\Delta = 0.5 $
and $ M_\Delta = 10^{-4} H_{\rm inf} \gg m_{3/2} $.
The dots represent the times of
$ t/t_0 = 1, 10 , 10^2 , 10^3 , 10^4 , 10^5 $.}
\label{tra}
\end{center}
\end{figure}

\newpage

\end{document}